\documentclass[12pt]{article}
\usepackage{graphicx}
\begin{document}

\centerline{\bf Simulation and Experiment of Extinction or Adaptation}
\medskip

\centerline{\bf of Biological Species after Temperature Changes}
\bigskip

\centerline{D. Stauffer$^1$ and H. Arndt$^2$}
\bigskip

\centerline{Cologne University, D-50923 K\"oln, Euroland}

\bigskip
$^1$ Institute for Theoretical Physics, e-mail: stauffer@thp.uni-koeln.de

$^2$ Zoological Institute, e-mail: hartmut.arndt@uni-koeln.de\\

\begin{abstract}
Can unicellular organisms survive a drastic temperature change, and adapt to it
after many generations? In simulations of the Penna model of biological ageing,
both extinction and adaptation were found for asexual and sexual reproduction
as well as for parasex. These model investigations are the basis for the design
of evolution experiments with heterotrophic flagellates.
\end{abstract}

\medskip

Global warming may be a relatively slow process, but the drastic temperature
drop after a full-scale nuclear war (``nuclear winter'') or its more speculative
form shown in the Hollywood movie ``The day after tomorrow'' may cause the
extinction of mankind. Though global warming may slowly change the mean 
temperature, it is expected that amplitudes of temperature fluctuations 
increase significantly. Example for adaptations to unusual temperatures are 
thermophilic bacteria near hot springs, or simple eukaryotes having adjusted to 
the higher temperatures in high-temperature sewage treatment plants. We plan 
biological 
experiments with heterotrophic flagellates (diploid and unicellular) to check 
how these protists survive a rapid or slow temperature change, by adaptation
to the new environment over many generations. To guide these experiments which
may take years we report here much simpler computer simulations of the same
effect.

The Penna model \cite{penna,book} is the most often used computer model for 
biological ageing. It represents the DNA genome by a computer word of, say,
32 bits such that bit = zero means a favourable (``wild type'') allele
while bit = 1 represents a deleterious heritable mutation, starting to affect 
the health from  that age on to which the bit position corresponds. Thus the 
whole lifespan is divided into 32 time units, which may correspond to several 
years for humans and less than an hour for bacteria and protists. At each
time unit, each living adult beyond the minimum reproduction age of eight
time units produces asexually one offspring with probability $b$. The offspring 
has the same genome as the parent except for one detrimental mutation at a 
random bit position; if the bit is already mutated it stays as it is.
Three active bad mutations kill the individual. For sexual reproduction, each
individual has two bit-strings, produces gametes by crossover, and one male and 
one female gamete are combined to allow the mother to give birth. For parasex
of haploid bacteria \cite{parasex}, parts of the bit-strings of two individuals
are exchanged, but no distinction between male and female is made. Further
details including computer programs are given in \cite{book}, somewhat updated
in \cite{vancouver}. We use the parameters of the programs listed there 
except that the birth rate $b$ is reduced to 0.3. 

As in \cite{suzana}, a rapid change of the environment (here: temperature) is
simulated by assuming that at one certain bit position, what was before a good
allele (bit = 0) now becomes bad, and what was bad (bit = 1) now becomes good.
(This might represent the production or suppression e.g. of heat shock proteins.)
Technically for haploid genomes this is achieved by changing all the bits at 
this position. We simulated 20,000 time units, and changed the ``temperature"
at time step 10,000. The position of the bit which changes its meaning was
typically 12, slightly higher than the minimum reproduction age of 8.

\begin{figure}[hbt]
\begin{center}
\includegraphics[angle=-90,scale=0.4]{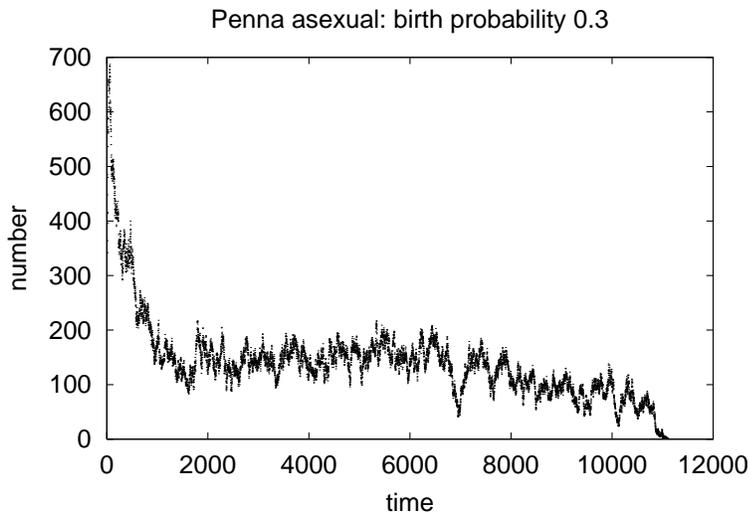}
\end{center}
\caption{Extinction of an asexual population shortly after a change at time =
10,000; birth rate 0.3.
}
\end{figure}

\begin{figure}[hbt]
\begin{center}
\includegraphics[angle=-90,scale=0.4]{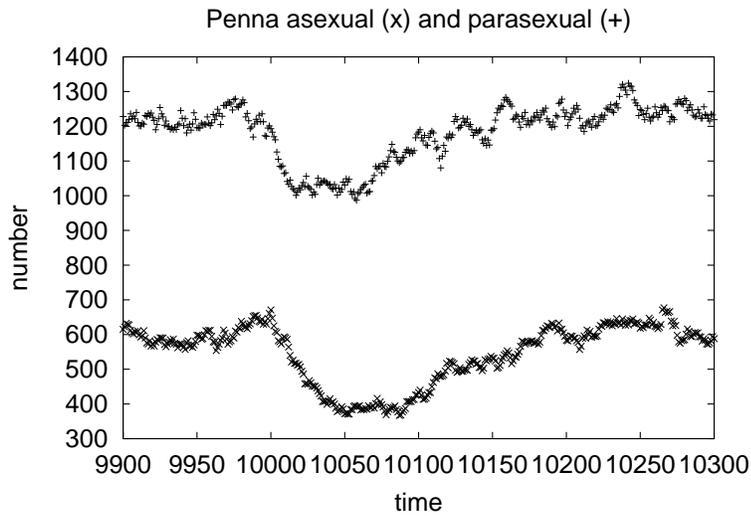}
\end{center}
\caption{Survival of a larger population (x); otherwise parameters as in Fig.1.
The upper (+) symbols refer to the more favourable parasex.
}
\end{figure}

\begin{figure}[hbt]
\begin{center}
\includegraphics[angle=-90,scale=0.4]{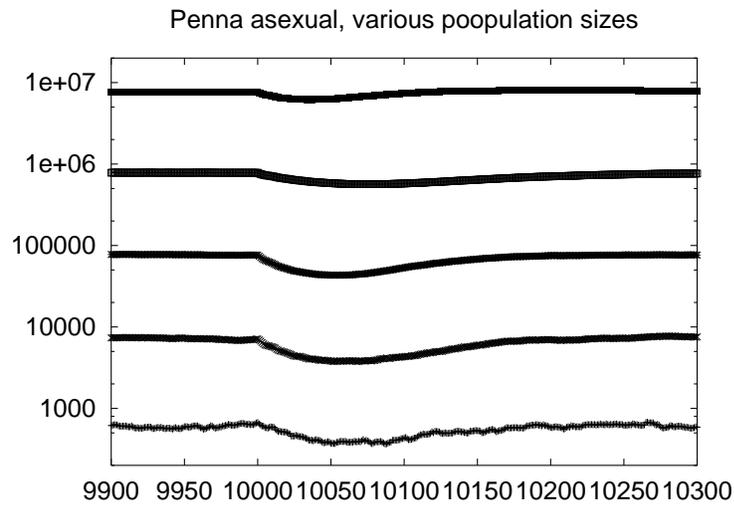}
\end{center}
\caption{Dependence of fluctuations and systematic change on population
size; otherwise parameters as in Figs. 1,2.
}
\end{figure}

\begin{figure}[hbt]
\begin{center}
\includegraphics[angle=-90,scale=0.4]{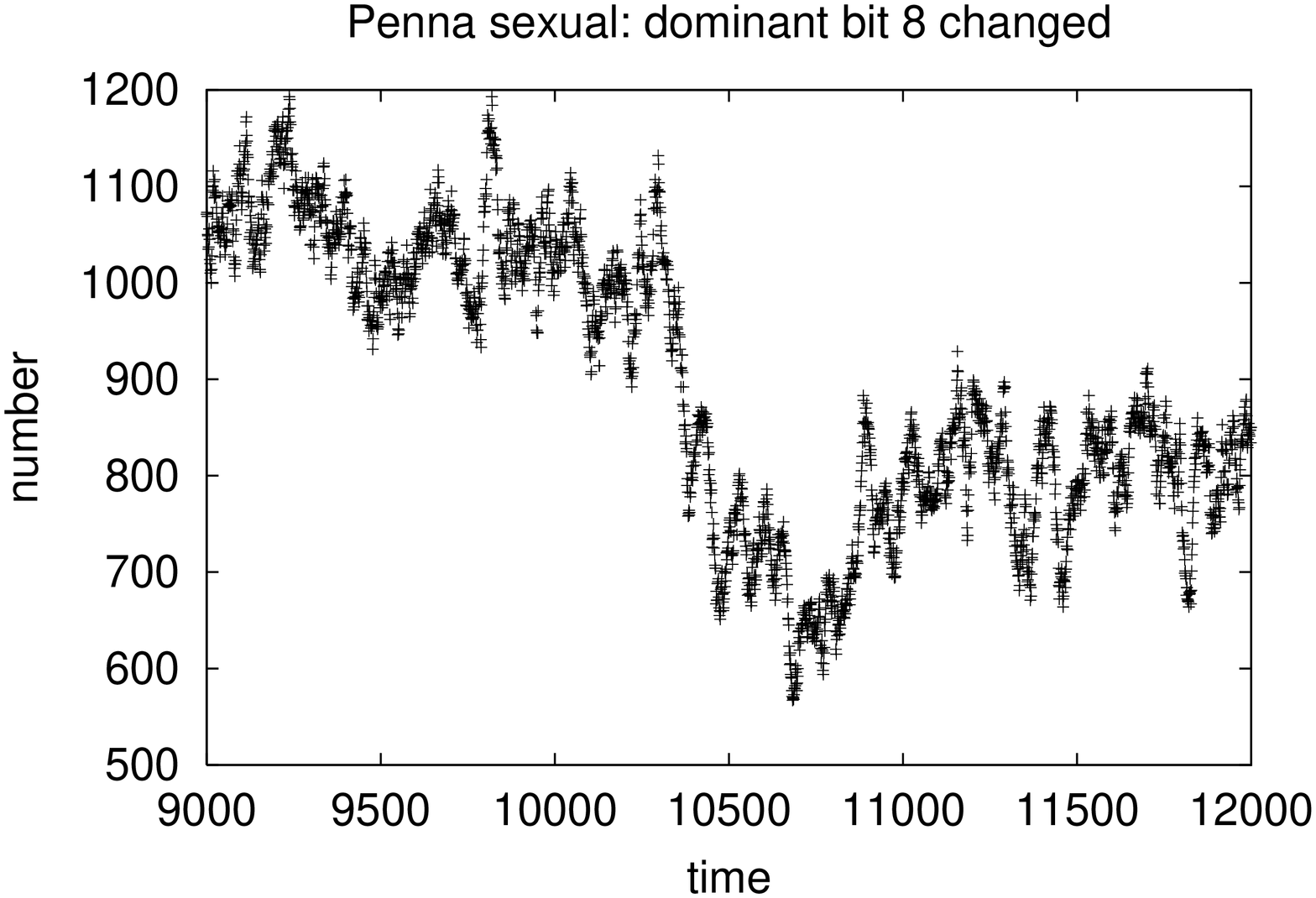}
\end{center}
\caption{Effect on a sexual population. Birth rate 0.6 for females. Note
expanded time interval shown. The changed bit was at position 8 and dominant.
}
\end{figure}

Fig. 1 shows that the temperature shock can lead to the extinction of the 
whole population \cite{suzana}. By starting with a larger population, this 
extinction can be avoided, Fig.2. Increasing the population further reduces
the relative fluctuations, Fig.3. Also with parasex, Fig.2, or sexual 
reproduction, Fig.4,
similar effects are observed. We see thus how the temperature
change at first reduces the survival, and the population diminishes
drastically. After some time, enough individuals who had the initially bad and
later good mutation have produced enough offspring so that
the total population recovers slowly.

Our biological experiments thus far are of the type of Fig.1; we still have to
search for conditions where not the whole population dies out.

In conclusion these simulations suggest that high population densities ($\gg
10^3$ individuals) and an experimental time consisting of hundreds of 
generations are necessary prerequisites to successfully study genetic adaptation
to changed temperatures. The computational analysis indicated that the study
of bacteria and nanoprotists (e.g. heterotrophic flagellates) should offer a 
suitable possibility for evolutionary experiments lasting only a few years.
 
\bigskip
{\bf Acknowledgements}: We thank M. L\"assig and D. Tautz for stimulating this 
cooperation.

\end{document}